\documentclass[showpacs,preprintnumbers,prd,nofootinbib,floats,amssymb,floatfix]{revtex4}
\usepackage{amsmath}
\usepackage{amsfonts}
\usepackage{graphicx}
\usepackage{amsxtra}
\usepackage{eufrak}
\usepackage{hyperref}
\usepackage{amsmath}
\usepackage{amstext}
 
\begin{document}
\title{ On the Faddeev-Jackiw symplectic framework for topologically massive gravity}
\author{Alberto Escalante}  \email{aescalan@ifuap.buap.mx}
 \affiliation{  Instituto de F{\'i}sica, Benem\'erita Universidad Aut\'onoma de Puebla, \\
 Apartado Postal J-48 72570, Puebla Pue., M\'exico, }
\author{ Omar Rodr{\'i}guez-Tzompantzi}  \email{omar.tz2701@gmail.com}
 \affiliation{ Facultad de Ciencias F\'{\i}sico Matem\'{a}ticas, Benem\'erita Universidad Au\-t\'o\-no\-ma de Puebla,
 Apartado postal 1152, 72001 Puebla, Pue., M\'exico.}
\begin{abstract}
The dynamical structure of topologically massive gravity  in the context of the Faddeev-Jackiw symplectic approach is studied.  It is shown that this method allows us to avoid some ambiguities arising in the study of the gauge structure via the Dirac formalism. In particular, the complete set of  constraints  and the  generators of the  gauge symmetry  of the theory are obtained straightforwardly via the zero-modes of the symplectic matrix. In order to obtain  the generalized  Faddeev-Jackiw  brackets and calculate the local physical degrees of freedom of this model, an appropriate gauge-fixing procedure is introduced. Finally, the similarities and advantages between  the Faddeev-Jackiw method and   Dirac's formalism  are briefly discussed.
\end{abstract}
 \date{\today}
\pacs{98.80.-k,98.80.Cq}
\preprint{}
\maketitle
\section{Introduction}
Theories of interacting spin-2 fields, such as  massive gravity, have been considerably studied in the literature, with particular focus on their symmetries and physical degrees of freedom \cite{deser1, deser2,Bergshoeff,Infraded, aspects, massive, 2-spin, lorentz-massive,Charges, Covariant}. The construction of a unitary and renormalizable theory of gravity with propagating degrees of freedom  has been a long-sought goal towards our understanding of gravitation. In this context, it is well known that the Fierz-Pauli theory  provides a consistent description of the linear fluctuations of a massive graviton on a flat space-time \cite{pauli}. The  Boulware-Deser (BD) ghost mode was exactly found for the Fierz-Pauli theory
taken at the non-linear level \cite{ghost, ghost2},  which violates the unitarity of the theory (a condition of consistency in quantum gravity). For this reason, the construction of an action for nonlinear massive gravity  must ensure the absence of any  ghost-like unphysical degrees of freedom, thereby rendering a  stable and consistent theory. Strictly speaking, the theory must possess the necessary dynamical constraints for removing  the ghost degrees of freedom, Nonetheless, it would be interesting if one could systematically obtain the constraints that eliminate such ghost fields. This approach would also be helpful to understand the gauge structure and the physical content of this kind of theories. In this work, we are interested in the study of the three-dimensional version of a massive gravity theory.  \\
It is well known  that  the key ingredient for understanding the physical content of a gauge dynamic system lies in the identification of the physical degrees of freedom along with observable quantities and symmetries.  Therefore, in a gauge theory  it is essential to make the distinction between gauge-invariant (gauge-dependent) quantities, which do (do not) correspond to observable quantities \cite{observables, benarjee}, though the former are not necessarily present at the quantum level. The task of identifying the  symmetries and observable quantities in a physical theory is, in general, non trivial, specially in gauge theories with general covariance, such as general relativity. Nevertheless, there are two approaches to obtain in a systematic way the symmetries and conserved quantities of a particular physical theory:  Dirac's formalism \cite{Dirac} and Faddeev-Jackiw [FJ] method \cite{Faddeev}. In the former approach, it is necessary to classify all constraints into first- and second-class ones. As  a consequence, the physical degrees of freedom can be exactly counted, and a generator of the gauge symmetry can be constructed  as a  suitable combination of the first-class constraints in order to identify the physical observables \cite{Castellani}. Furthermore, the brackets to quantize a gauge system (Dirac's brackets)   can be obtained  by getting rid of   the second-class constraints   \cite{teitelboim,Henneaux}. On the other hand, the F-J method provides a symplectic approach for constrained systems based on a first-order Lagrangian. The basic feature of this approach is that one it is not necessary to classify  the constraints into first- and second-class ones. Still, several essential elements of a physical theory, such as the degrees of freedom, the gauge symmetry and  the quantization brackets  (generalized  F-J brackets)   can also be obtained (see \cite{barcelos, barcelos2, pons,modified,bufalo} for a review). In this framework, the non-null F-J brackets emerge from the symplectic matrix. For a gauge system, this matrix remains singular unless a gauge-fixing procedure is introduced. In addition, the generators of the gauge symmetries are given in terms of the zero modes of this symplectic matrix. In this respect, the F-J symplectic method provides a straightforward  effective tool to deal with gauge theories because it is algebraically simpler than  Dirac's formalism. In particular, if   secondary,  tertiary, or higher order constraints  are present.

Quite recently, the  F-J symplectic method has proved  useful in the study of many physical theories, for instance in the construction of Maxwell-inspired $SU(3)$-like and $SU(3)\otimes SU(2)\otimes U(1)$ non-Abelian theories \cite{qcd}, as well as  noncommutative gauge theories \cite{noncommutativity}. Furthermore, this approach not only has been  useful to study non-Abelian systems \cite{non-abelian}, hidden symmetries \cite{hidden} and self-dual fields \cite{selfdual}, but also to quantize massive non-Abelian Yang-Mills fields, and to study the extended Horava-Lifshitz gravity \cite{Horava}. For other works on the F-J symplectic approach we refer the interested reader to Ref. \cite{exotica, stukel, omaryo}.

The purpose of the present work is to present a detailed study of three-dimensional topologically massive gravity (TMG) in a completely different context to that presented in \cite{Daniel, Mu, blagojevic, Carlip}. It is well-know that the canonical analysis of TMG is a  large and tedious  task since there are present secondary, tertiary   and quartic constraints with a complicated algebra \cite{blagojevic, Carlip}. On the other hand, it is possible  that  if  one step of the Dirac's formalism is either incorrectly  applied  or  omitted  \cite{5, 6},   the results could be incorrect \cite{Mu, Carlip}. In this respect, we will apply the  F-J symplectic approach to systematically obtain the constraints necessary to remove the unphysical degree of freedom of the theory, the gauge symmetries,  and the fundamental  F-J brackets  by introducing an appropriate gauge-fixing procedure.  Moreover, the similarities and advantages between this procedure and Dirac formalism will be discussed. It also will be shown that the physical degrees of freedom, the gauge symmetries and the brackets to quantize agree with those found via the Dirac method  in \cite{Mu, blagojevic, Carlip}.

The rest of the paper has been organized as follows. In Sec. II, we show that  the F-J symplectic method applied to TMG leads to an alternative way  for identifying the  dynamical constraints. In Sec. III, the gauge transformations are obtained using the zero-modes of the symplectic 2-form matrix. In Sec. IV, we show that both the fundamental F-J brackets and the physical degrees of freedom are obtained by introducing a gauge-fixing procedure.  In Sec. V, we present a  summary and  the conclusions.
%%%%%%%%%%%%%%%%%%%%%%%%%%%%%%%%%%
\section{Faddeev-Jackiw symplectic approach to TMG}
The action for TMG can be written as \cite{ Mu, blagojevic, Carlip}
\begin{eqnarray}
S[A,e,\lambda]&=&\int_{\mathcal{M}}\left[2\theta e^{i}\wedge F[A]_{i}+\lambda^{i}\wedge T_{i}+\frac{\theta}{\mu}A^{i}\wedge \left(dA_{i}+\frac{1}{3}f_{ijk}A^{j}\wedge A^{k}\right)\right],
\label{1}
\end{eqnarray}
where $\mu$ is the Chern-Simons parameter, $\theta = {1}/{16πG}$, and $A^{i}=A_{\mu}{^{i}}dx^{\mu}$ is a connection 1-form valued on the adjoint representation of the Lie group $SO(2,1)$, which admits an invariant totally anti-symmetric tensor $f_{ijk}$. Furthermore      $e^{i}=e_{\mu}^{i}dx^{\mu}$ is a triad 1-form that represents the  gravitational field and  $F^{i}$ is the curvature 2-form of the connection $A^{i}$, i.e.,  $F_{i}\equiv dA_{i}+\frac{1}{2}f_{ijk}A^{j}\wedge A^{k}$. Finally $\lambda^{i}$ is a Lagrange multiplier 1-form that ensures that  the torsion vanishes $T_{i}\equiv de_{i}+f_{ijk}A^{j}\wedge e^{k}=0$.\\
The equations of motion that arise from the variation of the action (\ref{1}) with respect to the dynamical variables $e_{\alpha}{}^{i}$, $A_{\alpha}{^{i}}$  and $\lambda_{\alpha}{^i{}}$ are given, in addition to some total derivative terms, by
\begin{eqnarray}
\left(\delta e\right)^{\alpha i}&=&\epsilon^{\alpha\nu\rho}\left(2\theta F_{\nu\rho}{^{i}}+D_{\nu}\lambda_{\rho}{^{i}}\right)=0,\nonumber\\
\left(\delta A\right)^{\alpha i}&=&\epsilon^{\alpha\nu\rho}\left(2\theta T_{\nu\rho}{^{i}}+f^{i}{_{jk}}\lambda_{\nu}{^{j}}e_{\rho}{^{k}}+2\theta\mu^{-1}F_{\nu\rho}{^{i}}\right)=0,\nonumber\\
\left(\delta \lambda\right)^{\alpha i}&=&\epsilon^{\alpha\nu\rho}T_{\nu\rho}{^{i}}=0.\label{eqm}
\end{eqnarray}
From the second and third equation in (\ref{eqm}), the Lagrange multiplier $\lambda_{\mu}{^{i}}$ can be solved in terms of the Schouten tensor of the manifold $\mathcal{M}$
\begin{equation}
\lambda_{\mu}{^{i}}=2\theta\mu^{-1}S_{\mu\nu}e^{i\nu},\hphantom{111}\mathrm{with}\hphantom{111}S_{\mu\nu}=(Ric)_{\mu\nu}-\frac{1}{4}g_{\mu\nu}R\label{mul}.
\end{equation}
The manifold  is endowed with a space-time metric, $g_{\mu\nu}=e_{\mu}{^{i}}e_{\nu}{^{j}}\eta_{ij}$. Furthermore, since the torsion vanishes, the spin-connection $A_{\mu}{^{i}}$ is a function of the dreibein $e_{\mu}{^{i}}$
\begin{equation}
 A_{\mu}{^{ij}}=-e^{\nu j}\partial_{\mu}e_{\nu}{^{i}}+\Gamma_{\alpha\mu}^{\beta}e_{\beta}^{i}e^{\alpha j},
\end{equation}
where $\Gamma_{\alpha\mu}^{\beta}$ is the Christoffel symbol. Inserting  the relation (\ref{mul}) into  the first equation of (\ref{eqm}, one gets the usual field equation of TMG \cite{deser1, Daniel} in the second-order formalism
\begin{equation}
G_{\mu\nu}+\frac{1}{\mu}C_{\mu\nu}=0,\label{second-eq}
\end{equation}
where $G_{\mu\nu} \equiv R_{\mu\nu}-\frac{1}{2}g_{\mu\nu}R$ is the Einstein tensor, and $C_{\mu\nu} \equiv \epsilon_{\mu}{^{\alpha\beta}}\nabla_{\alpha}S_{\beta\nu}$ is essentially the (symmetric traceless ) Cotton tensor obtained from varying the gravitational Chern-Simons term with respect to the metric, with $\nabla=\partial+\Gamma$  the covariant derivative for the Christoffel connection. In addition, the particle content of this theory can be seen by performing a linearized approximation to the field equations about a Minkowski background \cite{deser1} (see appendix A).\\
In order to perform the symplectic analysis, we will assume that the manifold $\mathcal{M}$ is topologically $\Sigma\times \Re$, where $\Sigma$ corresponds to a Cauchy's surface without  boundary $(\partial\Sigma = 0)$ and $\Re$ represents an evolution parameter. Here, $x^{\mu}$ are the coordinates that label the points of the 3-dimensional manifold $\mathcal{M}$. In our notation, Greek letters run from 0 to 2, while the middle alphabet letters $(i,j,k,...)$ run from $1$ to $3$.\\
By performing the $2 + 1$ decomposition of our fields without breaking the internal  symmetry, we can write the action (\ref{1}) as
\begin{eqnarray}
S[A,e,\lambda]&=&\int_{\Sigma}\left[\theta\epsilon^{ab}\left(2e_{bi}\ + \frac{1}{\mu}A_{bi}\right)\dot{A}^{i}{_{a}} + \epsilon^{ab}\lambda_{ib}\dot{e}^{i}{_{a}} + \epsilon^{ab}e^{i}{_{0}}\left(\theta F_{abi} + D_{a}\lambda_{bi}\right) + \frac{1}{2}\epsilon^{ab}\lambda^{i}{_{0}}T_{abi}\right.\nonumber \\
&&+\left. \epsilon^{ab}A^{i}{_{0}}\left(\theta T_{abi} + \frac{1}{\mu}\theta F_{abi} +f_{ijk}\lambda^{j}{_{a}}e^{k}{_{b}}\right)\right]d^{3}x,
\label{2}
\end{eqnarray}
where $F_{ab}{^{i}}=\partial_{a}A_{b}{^{i}}-\partial_{b}A_{a}{^{i}}+f^{i}{_{jk}}A_{a}{^{j}}A_{b}{^{k}}$, $T_{ab}{^{i}}=D_{a}e_{b}{^{i}}-D_{b}e_{a}{^{i}}$, and the covariant derivative of $\lambda_{a}{^{i}}$ is defined as $D_{a}\lambda_{b}{^{i}}=\partial_{a}\lambda_{b}{^{i}}+f^{i}{_{jk}}A_{a}{^{j}}\lambda_{b}{^{k}}$. Here $a,b = 1,2$ are space coordinate indices (the dot represents a derivative with respect to the evolution parameter). From (\ref{2}) we can identify the following first-order Lagrangian density
\begin{eqnarray}
\mathcal{L}^{(0)}&=&\theta\epsilon^{ab}\left(2e_{bi} + \frac{1}{\mu}A_{bi}\right)\dot{A}^{i}{_{a}}+ \epsilon^{ab}\lambda_{ib}\dot{e}^{i}{_{a}} + \epsilon^{ab}e^{i}{_{0}}(\theta F_{abi} + D_{a}\lambda_{bi}) + \frac{1}{2}\epsilon^{ab}\lambda^{i}{_{0}}T_{abi}\nonumber \\
&&+\epsilon^{ab}A^{i}{_{0}}\left(\theta T_{abi} + \frac{1}{\mu}\theta F_{abi} +f_{ijk}\lambda^{j}{_{a}}e^{k}{_{b}}\right).\label{3}
\end{eqnarray}
From the variational principle applied to the  Lagrangian density (\ref{3}), it is possible to write the symplectic equations of motion as
\begin{equation}
f_{ij}^{(0)}\dot{\xi}^{(0)j}=\frac{\delta V^{(0)}(\xi)}{\delta\xi^{(0)i}},
\label{motion}
\end{equation}
where $f_{ij}^{(0)}=\frac{\delta }{\delta\xi^{(0)i}}a_{j}^{(0)}(\xi)-\frac{\delta}{\delta\xi^{(0)j}}a_{i}^{(0)}(\xi)$, which is clearly antisymmetric, is known as the symplectic two-form, which yields the following symplectic variable set ${\xi}{^{(0)i}} = (A^{i}{_{a}}, A^{i}{_{0}}, e^{i}{_{a}},$
$e^{i}{_{0}}, \lambda^{i}{_{a}}, \lambda^{i}{_{0}})$, the corresponding symplectic 1-form ${a}^{(0)}{_{i}} = (2\theta\epsilon^{ab}e_{bi}+\frac{\theta}{\mu}\epsilon^{ab}A_{bi}, 0, \epsilon^{ab}\lambda_{bi}, 0,$
$ 0, 0)$, and the symplectic potential $V^{(0)}$ given by
\begin{eqnarray}
V^{(0)}=\epsilon^{ab}e^{i}{_{0}}(\theta F_{abi} + D_{a}\lambda_{bi}) + \frac{1}{2}\epsilon^{ab}\lambda^{i}{_{0}}T_{abi}+\epsilon^{ab}A^{i}{_{0}}\left(\theta T_{abi} + \frac{1}{\mu}\theta F_{abi} +f_{ijk}\lambda^{j}{_{a}}e^{k}{_{b}}\right).\label{potential}
\end{eqnarray}
By using the symplectic variables, we find that the symplectic matrix $f^{(0)}_{ij}$ can be written as
\begin{eqnarray}
\label{eq}
f_{ij}^{(0)}(x,y)=\left(
  \begin{array}{cccccc}
 2\frac{\theta}{\mu}\epsilon^{ab}\eta_{ij}    &	\quad   0   &\quad  -2\theta\epsilon^{ab}\eta_{ij}   &\quad  0     & \quad 0  &\quad  0 	 	 \\
 0 &\quad   0   &\quad  0   &\quad   0 &\quad   0   &\quad   0 \\
2\theta\epsilon^{ab}\eta_{ij}   &\quad  0   &\quad   0      &\quad   0   &\quad    -\epsilon^{ab}\eta_{ij}   &\quad  0 	\\
    0   &\quad  0   &\quad   0    &\quad 0  &\quad  0	 &\quad  0 \\
0  &\quad  0  &\quad    \epsilon^{ab}\eta_{ij}  &\quad  0  &\quad  0 	&\quad  0 	 \\
0   &\quad   0  &\quad  0  &\quad  0   &\quad  0 	&\quad  0
 \end{array}
\right)\delta^{2}(x-y).
\end{eqnarray}
Clearly  $f^{(0)}_{ij}$ is degenerate, which means that there are more degrees of freedom in the equations of motion (\ref{motion}) than physical degrees of freedom in the theory.  We thus have a constrained theory,  with constraints that must remove the unphysical degrees of freedom. The zero-modes of this matrix turn out to be $(v_{1}^{(0)})^{T}_{i}= (0,v^{A{^{i}{_{0}}}}, 0, 0, 0, 0)$, $(v_{2}^{(0)})^{T}_{i}=(0, 0, 0, v^{e{^{i}{_{0}}}}, 0, 0)$ and $(v_{3}^{(0)})^{T}_{i}=(0, 0, 0, 0, 0, v^{\lambda{^{i}{_{0}}}})$, where $v^{A{^{i}{_{0}}}},v^{e{^{i}{_{0}}}}$ and $v^{\lambda{^{i}{_{0}}}}$ are arbitrary functions. By multiplying the two sides of (\ref{motion}) by these zero-modes, we can obtain the following primary constraints
\begin{eqnarray}
\Xi^{(0)}_{i}&=& \int\ dx^{2}(v_{1}^{(0)})^{T}_{j}\frac{\delta}{\delta\xi^{(0)j}}\int\ dy^{2} V^{(0)}  = \theta\epsilon^{ab}T_{abi} + \frac{\theta}{\mu}\epsilon^{ab}F_{abi} + \epsilon^{ab} f_{ijk}\lambda^{j}{_{a}}e^{k}{_{b}}=  0, \nonumber\label{4.1}\\
\Theta^{(0)}_{i}&=& \int\ dx^{2}(v_{2}^{(0)})^{T}_{j}\frac{\delta}{\delta\xi^{(0)j}}\int\ dy^{2}V^{(0)} = \theta\epsilon^{ab}F_{abi} + \epsilon^{ab}D_{a}\lambda_{bi} = 0,\nonumber \label{4.2}\\
\Sigma^{(0)}_{i}&=& \int\ dx^{2}(v_{3}^{(0)})^{T}_{j}\frac{\delta}{\delta\xi^{(0)j}}\int\ dy^{2}V^{(0)} =  \frac{1}{2}\epsilon^{ab} T_{abi} = 0.
\label{4.3}
\end{eqnarray}
Following the prescription of the  symplectic formalism,  we will analyze wheter there are new constraints. For this aim, we impose a consistency condition on the constraints (\ref{4.3}) as in the Dirac method:
\begin{equation}
\dot{\Omega}^{(0)}=\frac{\delta\Omega^{(0)}}{\delta\xi^{(0)i}}\dot{\xi}^{(0)i}=0\hphantom{111}\mathrm{with}\hphantom{111}\Omega^{(0)}=\Xi_{i}^{(0)},\Theta_{i}^{(0)}, \Sigma_{i}^{(0)},\label{consistency}
\end{equation}
which means that these constraints must be preserved in time.  The consistency condition on the primary constraints (\ref{consistency}) and (\ref{motion}) can be rewritten as
\begin{eqnarray}
f_{kj}^{(1)}\dot{\xi}^{(0)j}=Z_{k}^{(1)}(\xi),
\label{p}
\end{eqnarray}
where
\begin{eqnarray}
f_{kj}^{(1)}=\left(
\begin{array}{cc}
f^{(0)}_{ij} \\
\frac{\delta\Omega^{(0)}}{\delta\xi^{(0)j}}
\end{array}\right)\hphantom{111}\mathrm{and}\hphantom{111}Z_{k}^{(1)}=
\left(
\begin{array}{cc}
\frac{\delta V^{(0)}}{\delta\xi^{(0)j}} \\
0\\
0\\
0
\end{array}
\right).
\end{eqnarray}
Thus the new symplectic matrix $f_{kj}^{(1)}$ is given by
\begin{eqnarray}
\label{eq}
&&
\left(
 \begin{array}{cccccc}
 2\frac{\theta}{\mu}\eta_{ij}    &  0   &  -2\theta\eta_{ij}   &0     & 0  &  0 	 	 \\
 0 &  0   & 0   &  0 &  0   &   0 \\
2\theta\eta_{ij}   &  0   & 0      &  0   &   -\eta_{ij}   &  0 	\\
 0   &  0   &  0    & 0  &  0	 & 0 \\
0  &  0 & \eta_{ij}   & 0 	& 0 &0	 \\
0   &   0  & 0  & 0   &  0 	&  0 \\
2\frac{\theta}{\mu}(\eta_{ij}\partial_{a}-f_{ijk}A^{k}{_{a}} -\mu f_{ijk}e^{k}{_{a}})	&	0	&
2\theta(\eta_{ij}\partial_{a}-f_{ijk}A^{k}{_{a}}-\frac{1}{2\theta}f_{ijk}\lambda^{k}{_{a}})	&	0	&	-f_{ijk}e^{k}{_{a}}	
&	0	\\
2\theta(\eta_{ij}\partial_{a}-f_{ijk}A^{k}{_{a}}-\frac{1}{2\theta}f_{ijk}\lambda^{k}{_{a}})	&	0	&	0	&	0	&	(\eta_{ij}\partial_{a}-f_{ijk}A^{k}{_{a}})	&	0\\
-f_{ijk}e^{k}{_{a}}	&	0	&	(\eta_{ij}\partial_{a} - f_{ijk}A^{k}{_{a}})		&	0	&	0	&	0
\end{array} 		
\right)\nonumber\\
&&\times\epsilon^{ab} \delta^{2}(x-y).
\label{6}
\end{eqnarray}
Although $f_{kj}^{(1)}$ is not a square matrix, it still has the following linearly independent zero-modes
\begin{eqnarray}
(v_{1}^{(1)}){^{j}}{^{T}}&=& \left(\partial_{a}v^{j}-f^{j}{_{lm}}A^{l}{_{a}}v^{m}, v^{e{^{j}{_{0}}}}, -f^{j}{_{lm}}e^{l}{_{a}}v^{m}, 0, f^{j}{_{lm}}\lambda_{a}{^{l}}v^{m}, 0, v^{j}, 0, 0\right),\nonumber \\
(v_{2}^{(1)}){^{j}}{^{T}}&=& \left(-\frac{\mu}{2\theta}f^{j}{_{lm}}e^{l}{_{a}}v^{m}, 0, 0, v^{A{^{j}{_{0}}}}, \partial_{a}v^{j} - f^{j}{_{lm}}A^{l}{_{a}}v^{m} - \mu f^{j}{_{lm}}e^{l}{_{a}}v^{m}, 0, 0, 0, v^{j}\right),\nonumber \\
(v_{3}^{(1)}){^{j}}{^{T}}&=& \left(-\frac{\mu}{2\theta}f^{j}{_{lm}}\lambda^{l}{_{a}}v^{m}, 0, \partial_{a}v^{j} - f^{j}{_{lm}}A^{l}{_{a}}v^{m}, 0, -\mu f^{j}{_{lm}}\lambda^{l}{_{a}}v^{m}, v^{\lambda{^{j}{_{0}}}}, 0, v^{j}, 0\right),
\label{7}
\end{eqnarray}
where $v^{m}, v^{e_0^{j}}, v^{A_0^{j}}, v^{\lambda ^{j}_0}$ are arbitrary functions. On the other hand, the matrix $Z_{k}^{(1)}$ is given by
\begin{eqnarray}
\label{eq}
\left(
  \begin{array}{c}
-  2\theta D_{a}e{_{0j}}+f{_{jlm}}e_{0}{^{l}}\lambda_{a}{^{m}}+f{_{jlm}}\lambda_{0}{^{l}}e_{a}{^{m}}+2\theta f{_{jlm}}A_{0}{^{l}}e_{a}{^{m}}-2\frac{1}{\mu} \theta D_{a}A{_{0j}}\\
0 \\
-D_{a}\lambda{_{0j}}-2\theta D_{a}A{_{0j}} + f{_{jlm}}A^{l}{_{0}}\lambda^{m}{_{a}} \\
0 \\
-D_{a}e{_{0j}}+ f{_{jlm}}A^{l}{_{0}}e^{m}{_{a}}\\
0 \\
0 \\
0 \\
0 \\
\end{array}
\right)\epsilon^{ab} \delta^{2}(x-y).\nonumber\\
\label{8}
\end{eqnarray}
 By performing the contraction of the two sides of  (\ref{p}) with the zero-modes (\ref{7}),  we can obtain the following constraints
\begin{eqnarray}
(v^{(1)})^{T}_{k}Z_{k}^{(1)}\mid_{\Omega^{(0)}=0}=0.
\label{9}
\end{eqnarray}
The substitution $\Omega^{(0)}=0$ guarantees that these constraints will drop from the remainder of the calculation. After a lengthy but straightforward calculation,  from (\ref{9}) we obtain the explicit form of the  secondary constraints
\begin{eqnarray}
\Lambda= 2\epsilon^{ab} e^{i}{_{a}}\lambda_{ib},\hphantom{111}\Lambda_{0a}= e^{i}{_{0}}\lambda_{ia} - e^{i}{_{a}}\lambda_{i0}.\label{10}
\label{new}
\end{eqnarray}
This agrees completely with what was found in \cite{blagojevic} by using the Dirac approach, however, in this formalism the constraints (\ref{10}) arise as tertiary and quartic constraints, respectively. Furthermore, we can impose the consistency conditions on (\ref{new}) to obtain the following system
\begin{eqnarray}
f_{kj}^{(2)}\dot{\xi}^{(0)j}=Z_{k}^{(2)}(\xi),
\end{eqnarray}
where we have now
\begin{eqnarray}
f_{kj}^{(2)}=\left(
\begin{array}{cc}
f^{(1)}_{ij} \\
\frac{\delta\Omega^{(1)}}{\delta\xi^{(0)j}}
\end{array}\right),\hphantom{111}\Omega^{(1)}=\Lambda, \Lambda_{0a}\hphantom{111}\mathrm{and}\hphantom{111}Z_{k}^{(2)}=
\left(
\begin{array}{cc}
Z_{k}^{(1)} \\
0\\
0
\end{array}
\right).
\end{eqnarray}
It is easy to see that, even after calculating the symplectic matrix $f_{kj}^{(2)}$ and  inserting the above constraints (\ref{10}), the zero-modes do not yield new constraints, which means that there are no further constraints in the theory, and thus our procedure  comes to an end.  We can now  introduce the Lagrange multipliers for the constraints (\ref{4.3}) and (\ref{10}) into the Lagrangian  density (\ref{3}) in order to construct a new one
\begin{eqnarray}
{\mathcal{L}}^{(3)}&=& \theta\epsilon^{ab}\left(2e^{i}{_{b}}+\frac{1}{\mu}A_{bi}\right)\dot{A}{^{i}{_{a}}} + \epsilon^{ab}\lambda_{ib}\dot{e}{^{i}{_{a}}} - \epsilon^{ab}(\theta F_{abi} + D_{a}\lambda_{bi})\dot{\alpha}^{i}-\frac{1}{2}\epsilon^{ab}T_{abi}\dot{\Gamma}^{i}\nonumber\\
&&- \epsilon^{ab}(\theta T_{abi} + \frac{\theta}{\mu}F_{abi} + f_{ijk}\lambda^{j}{_{a}}e^{k}{_{b}})\dot{\beta}^{i}  -\Lambda\dot{\varphi} - \Lambda_{0a}\dot{\varphi}^{0a},
\end{eqnarray}
where the new symplectic potential $V^{(3)}$ vanishes since it is a linear combination of constraints reflecting the general covariance of the theory, namely, $V^{(3)}=V^{(0)}\mid_{\Omega^{(0)}, \Omega^{(1)}=0}=0$. On the other hand, the new Lagrange multipliers enforcing the constraints are  $\dot{\alpha}{^{i}}=e_{0}^{i}$, $\dot{\beta}{^{i}}=A_{0}^{i}$, $\dot{\Gamma}{^{i}}=\lambda_{0}^{i}$, $\dot{\varphi}$ and  $\dot{\varphi}^{0a}$. Therefore, the new symplectic variable set is taken as
\begin{equation}
\xi^{(3)i}=\left(A^{i}{_{a}}, \beta^{i}, e^{i}{_{a}},\alpha^{i}, \lambda^{i}{_{a}},\Gamma^{i},\varphi,\varphi^{0a}\right).\label{final-variables}
\end{equation}
Thus, the corresponding symplectic 1-form is
\begin{equation}
a^{(3)}_{i}= \left(\theta\epsilon^{ab}\left(2e_{bi} + \frac{1}{\mu}A_{bi}\right), -\Xi^{(0)}{_{i}}, \epsilon^{ab}\lambda_{bi}, -\Theta^{(0)}_{i}, 0, -\Sigma^{(0)}_{i},-\Lambda, -\Lambda_{0a}\right).
\end{equation}
By using these symplectic variables, an explicit calculation yields a singular symplectic matrix $f^{(3)}_{ij}=\frac{\delta }{\delta\xi^{(3)i}}a_{j}^{(3)}(\xi)-\frac{\delta}{\delta\xi^{(3)j}}a_{i}^{(3)}(\xi)$. However, we have shown that there are no more constraints, therefore, the theory must have a local gauge symmetry. The zero-modes of $f^{(3)}_{ij}$  turn out to be
\begin{eqnarray}
\left(v_{1}^{(3)}\right)^{iT}&=&\left(-\partial_{a}\zeta^i-f^{i}{_{jk}}A_{a}{^{j}}\zeta^{k}, \zeta^{i},-f^{i}{_{jk}}e_{a}^{j}\zeta^{k},0,-f^{i}{_{jk}}\lambda_{a}{^{j}}\zeta^{k},0,0,0\right),\nonumber\label{v1}\\
\left(v_{2}^{(3)}\right)^{iT}&=&\left(-\frac{\mu}{2\theta}f^{i}{_{jk}}\lambda_{a}{^{j}}\kappa^{k}, 0,-\partial_{a}\kappa^i-f^{i}{_{jk}}A_{a}{^{j}}\kappa^{k},\kappa^{i},+\mu f^{i}{_{jk}}\lambda_{a}{^{j}}\kappa^{k},0,0,0\right),\nonumber\label{v2}\\
\left(v_{3}^{(3)}\right)^{iT}&=&\left(-\frac{\mu}{2\theta}f^{i}{_{jk}}e_{a}{^{j}}\varsigma^{k}, 0, 0, 0,-\partial_{a}\varsigma^i-f^{i}{_{jk}}A_{a}{^{j}}\varsigma^{k}+\mu f^{i}{_{jk}}e_{a}{^{j}}\varsigma^{k},\varsigma^{i},0,0\right)\label{v3}.
\end{eqnarray}
\section{Gauge symmetry}
It is well-known that the gauge symmetry determines the physical content of any gauge theory, therefore we need to know explicitly the fundamental gauge transformations of the theory. In agreement with the prescription of the symplectic formalism \cite{barcelos2, Montani, Mon-Woc, hidden}, the zero-modes correspond  to the generators of the gauge symmetry of the original theory (\ref{1}) (see Appendix B), i.e.
\begin{equation}
\delta_{G}\xi^{(3)i}=\left(v_{l}^{(3)}\right)^{iT}\epsilon^{l},
\end{equation}
where  $\{\left(v_{l}^{(3)}\right)^{iT}\}$ is the whole set of zero-modes of singular symplectic matrix $f_{ij}^{(3)}$ and $\epsilon^{l}$ stand for arbitrary infinitesimal parameters. By using this fact, the generators (\ref{v3}) yield the following fundamental gauge transformations of the basic fields
\begin{eqnarray}
\delta_{G} A_{\alpha}{^{i}}(x)&=&-D_{\alpha}\zeta^{i}-\frac{\mu}{2\theta}f^{i}{_{jk}}\left(e_{\alpha}{^{j}}\varsigma^{k}+\lambda_{\alpha}{^{j}}\kappa^{k}\right),\nonumber\label{g1}\\
\delta_{G} e_{\alpha}{^{i}}(x)&=&-D_{\alpha}\kappa^{i}-f^{i}{_{jk}}e_{a}^{j}\zeta^{k},\nonumber\label{g2}\\
\delta_{G} \lambda_{\alpha}{^{i}}(x)&=&-D_{\alpha}\varsigma^{i}-f^{i}{_{jk}}\lambda_{\alpha}{^{j}}\zeta^{k}+\mu f^{i}{_{jk}}\left(\lambda_{a}{^{j}}\kappa^{k}+e_{a}{^{j}}\varsigma^{k}\right)\label{g3},
\end{eqnarray}
where $\zeta^{i}$, $\kappa^{i}$ and $\varsigma^{i}$ are the  time-dependent gauge parameters. It is worth remarking that (\ref{g3}) correspond to the gauge symmetry of the theory, but not to diffeomorphisms. Nevertheless, it is known that an appropriate choice of the gauge parameters does, indeed, generate diffeomorphism (on-shell) \cite{jhep, Carlip, blagojevic2}. Thus, we can redefine the gauge parameters as
\begin{equation}
\zeta^{i}=-A^{i}{_{\mu}}\varepsilon^{\mu}, \hphantom{111}\kappa^{i}=-e^{i}{_{\mu}}\varepsilon^{\mu},\hphantom{111}\varsigma^{i}=-\lambda^{i}{_{\mu}}\varepsilon^{\mu},\label{lie}
\end{equation}
where $\varepsilon^{\mu}$ is an arbitrary three-vector.  In this manner from the fundamental gauge symmetry (\ref{g3}) and the mapping (\ref{lie}), we obtain
\begin{eqnarray}
\delta_{G}e_{\alpha}{^{i}} &=&\mathfrak{L}_{\varepsilon}e_{\alpha}{^{i}}-\varepsilon^{\mu}\epsilon_{\alpha\mu\nu}\left(\delta \lambda\right)^{\nu i},\nonumber\\
\delta_{G}A_{\alpha}{^{i}} &=&\mathfrak{L}_{\varepsilon}A_{\alpha}{^{i}}+\mu\varepsilon^{\mu}\epsilon_{\alpha\mu\nu}\left[\frac{1}{2\theta}\left(\delta A\right)^{\nu i}+\left(\delta \lambda\right)^{\nu i}\right],\nonumber\\
\delta_{G}\lambda_{\alpha}{^{i}} &=&\mathfrak{L}_{\varepsilon}\lambda_{\alpha}{^{i}}+2\mu\theta\varepsilon^{\mu}\epsilon_{\alpha\mu\nu}\left[\frac{1}{2\mu\theta}\left(\delta e \right)^{\nu i}-\frac{1}{2\theta}\left(\delta A\right)^{\nu i}+\left(\delta \lambda\right)^{\nu i}\right].
\end{eqnarray}
which are (on-shell) diffeomorphisms. On the other hand, diffeomorphism invariant theories have the Poincar\'e  transformations, as off-shell symmetries, by construction\cite{kibble, utiyama}. Thus, in order to recover the Poincar\'e  symmetries, we need to map the gauge parameters of fundamental gauge symmetries `$\delta_{G}$' (\ref{g3}) into those of the  Poincar\'e symmetries. This is achieved through the field-dependent map between the  gauge parameters (\ref{g3}) and the Poincar\'e ones \cite{blagojevic2}:
\begin{equation}
\zeta^{i}=A^{i}{_{\mu}}\varepsilon^{\mu}+\omega^{i}, \hphantom{111}\kappa^{i}=e^{i}{_{\mu}}\varepsilon^{\mu},\hphantom{111}\varsigma^{i}=\lambda^{i}{_{\mu}}\varepsilon^{\mu},
\end{equation}
where $\varepsilon^{\mu}$ and $\omega^{i}$ are the parameters of translations and local Lorentz rotations, respectively, which together constitute the $6$ gauge parameters of Poincar\'e symmetries in 3D. By using this map, it is  seen that the gauge symmetries indeed reproduce the Poincar\'e symmetries, but modulo terms proportional to the equations of motion
\begin{eqnarray}
\delta_{G}e_{\alpha}{^{i}} &=&-\varepsilon^{\mu}\partial_{\mu}e_{\alpha}{^{i}}-e_{\mu}{^{i}}\partial_{\alpha}\varepsilon^{\mu}-f^{i}{_{jk}}e_{\alpha}{^{j}}\omega^{k}+\varepsilon^{\gamma}\epsilon_{\alpha\gamma\nu}\left(\delta \lambda\right)^{\nu i},\nonumber\\
\delta_{G}A_{\alpha}{^{i}} &=&-\partial_{\alpha}\omega^{i}-f^{i}{_{jk}}A_{\alpha}{^{j}}\omega^{k}-\varepsilon^{\mu}\partial_{\mu}A_{\alpha}{^{i}}-A_{\mu}{^{i}}\partial_{\alpha}\varepsilon^{\mu}-\mu\varepsilon^{\gamma}\epsilon_{\alpha\gamma\nu}\left[\frac{1}{2\theta}\left(\delta A\right)^{\nu i}+\left(\delta \lambda\right)^{\nu i}\right],\nonumber\\
\delta_{G}\lambda_{\alpha}{^{i}} &=&-\varepsilon^{\mu}\partial_{\mu}\lambda_{\alpha}{^{i}}-\lambda_{\mu}{^{i}}\partial_{\alpha}\varepsilon^{\mu}-f^{i}{_{jk}}\lambda_{\alpha}{^{j}}\omega^{k}-2\mu\theta\varepsilon^{\gamma}\epsilon_{\alpha\gamma\nu}\left[\frac{1}{2\mu\theta}\left(\delta e \right)^{\nu i}-\frac{1}{2\theta}\left(\delta A\right)^{\nu i}+\left(\delta \lambda\right)^{\nu i}\right],\label{PGT}
\end{eqnarray}
where the equations of motion $\left(\delta e\right)^{\nu i}$, $\left(\delta A\right)^{\nu i}$ and $\left(\delta\lambda\right)^{\nu i}$ are defined in (\ref{eqm}). We thus conclude that the Poincar\'e symmetry (\ref{PGT}) as well as the diffeomorphisms (\ref{lie}) are contained in the fundamental gauge symmetry (\ref{g3}) only on-shell.  In addition, the generators of such gauge transformations can be represented in terms of the zero-modes, thereby making evident that the zero-modes of the symplectic two-form encode all the information about the gauge structure of this theory.
\section{Faddeev-Jackiw brackets}
Finally, in order to invert the symplectic matrix and obtain the generalized Faddeev-Jackiw  brackets and identify the physical degrees of freedom, we must introduce a gauge-fixing procedure, that is, new ``gauge constraints''. For convenience,  we use the temporal gauge, namely, $A^{i}{_{0}}=0$, $e^{i}{_{0}}=0$, $\lambda^{i}{_{0}}=0$ and $\varphi=cte$ (i.e. $\dot{\varphi}=0$).  As a direct consequence,  the term  $\Lambda_{0a}$ vanishes in the Lagrangian density. In this manner, we also introduce new Lagrange multipliers that enforce the gauge conditions, namely, $\rho_{i}$, $\omega_{i}$, $\tau_{i}$ and $\sigma$. Then, the final 1-form Lagrangian density reduces to
\begin{eqnarray}
{\mathcal{L}}{^{(4)}}&=&\theta\epsilon^{ab}\left(2e_{bi}+\frac{1}{\mu}A_{bi}\right)\dot{A}{^{i}{_{a}}}+\epsilon^{ab}\lambda_{ib}\dot{e}{^{i}{_{a}}}-\left(\Xi^{(0)}_{i} - \rho_{i}\right)\dot{\beta}{^{i}}-\left(\Theta^{(0)}_{i}-\omega_{i}\right)\dot{\alpha}{^{i}}\nonumber \\
&&-\left(\Sigma^{(0)}_{i} - \tau_{i}\right)\dot{\Gamma}{^{i}}-\left(\Lambda - \sigma\right)\dot{\varphi}.
\end{eqnarray}
Thus, we can identify the final symplectic variable set
\begin{equation}
\xi^{(4)i} = (A^{i}{_{a}},\beta^{i},e^{i}{_{a}},\alpha^{i},\lambda^{i}{_{a}},\Gamma^{i},\varphi,\rho_{i}, \omega_{i},\tau_{i},\sigma),
\label{variables}
\end{equation}
with the corresponding symplectic 1-form
\begin{equation}
{a}^{(4)}_{i} = \left(\theta\epsilon^{ab}\left(2 e_{bi} + \frac{1}{\mu}A_{bi}\right), -\Xi^{(0)}_{i} + \rho_{i}, \epsilon^{ab}\lambda_{bi}, -\Theta^{(0)}_{i} + \omega_{i}, 0, -\Sigma^{(0)}_{i}+ \tau_{i}, -\Lambda + \sigma, 0, 0, 0, 0\right).
\end{equation}
After some algebra, we obtain the explicit form of the symplectic two-form $f^{(4)}_{ij}$
\begin{eqnarray}
&&
{\small{}
\left(
\begin{array}{cccccccccccc}
\frac{2\theta}{\mu}F&-2\frac{\theta}{\mu}(A+\mu C)&-2\theta F&-2\theta( A+\frac{D}{2\theta})&0&-C&0&0&0&0&0\\
{2\frac{\theta}{\mu}}(A+\mu C)&0&2\theta( A+\frac{D}{2\theta})&0&C&0&0&-\eta_{ij}&0&0&0\\
2\theta F&-2\theta( A+\frac{D}{2\theta})&0&0&-F&-A&2I&0&0&0	&0\\
2\theta( A+\frac{D}{2\theta})&0&0&0&-A&0&0&0	&	-\eta_{ij}	&		0		&	0\\
0&-C&F&A&0&0&-2H		&0		&	0	&	0	&	0	\\
C&0&A&0&0&0&0	&	0	&	0	&	-\eta_{ij}	&	0	\\
0&0&-2I&0&2H&0&0	&	0	&	0	&	0	&	-1	\\
0&\eta_{ij}&0&0&0&0&0&0	&	0	&	0	&	0	\\
0&0&0&\eta_{ij}&0&0	&	0	&	0	&	0	&	 0&0	\\
0&0&0&0&0&\eta_{ij}&0	&	0	&	0	&	0	&	0&	\\
0&0&0&0&0&0&1	&	0	&	0	&	0	&	0	\\
\end{array}
\right)}\nonumber\\
&&\times\delta^{2}(x-y),
\end{eqnarray}
which is  non-singular and has the following inverse  $f^{(4)}{_{ij}}{^{-1}}$
\begin{eqnarray}
\left(
\begin{array}{cccccccccccc}
\frac{\mu}{2\theta}\overline{F}&0&0&0&-\mu\overline{F}&0&0&-\overline{A}&-\frac{\mu}{2\theta}\overline{D}&-\frac{\mu}{2\theta}\overline{C}&2\mu e_{b}{^{j}}\\
0&0&0&0&0&0&0&-\eta^{i}{_{j}}&0&0&0\\
0&0&0&0&-\overline{F}&0&0&-\overline{C} \overline{F}&-\overline{A}&0	&2e_{a}{^{l}}\\
0&0&0&0&0&0&0&0	&\eta^{i}_{j}&		0		&	0\\
\mu \overline{F}&0&\overline{F}&0&2\theta\mu \overline{F}&0&0&\overline{D}&\mu \overline{D}		&	2(\overline{A}-\mu \overline{C})	&	-2G\\
0&0&0&0&0&0&0	&	0	&	0	&	\eta^{i}{_{j}}	&	0	\\
0&0&0&0&0&0&0	&	0	&	0	&	0	&	1	\\
\overline{A}&\eta^{i}_{j}&C\overline{F}&0&\overline{D}&0&0&0	&	0	&	0	&	0	\\
\frac{\mu}{2\theta}D&0&A&-\eta^{i}_{j}&\mu \overline{D}&0	&	0	&	0	&	\frac{\mu}{ 2\theta}E	&	 0&0	\\
\frac{\mu}{2\theta}\overline{C}&0&0&0&(2\overline{A}-\mu \overline{C})&-\eta^{ij}&0	&	0	&	0	&	\frac{\mu}{\theta}B	&	2\mu \overline H&	\\
-2\mu e_{b}{^{i}}&0&-2e_{a}{^{l}}&0&2G&0&1	&	0	&	0	&	-2\mu \overline H	&	0\\
\end{array}
\right)\delta^{2}(x-y),\nonumber\\
\label{inv}
\end{eqnarray}
where
\[
A=\epsilon^{ab}\left(\partial_{a}\eta_{ij}+f_{ikj}A_{a}{^{k}}\right),\ C=\epsilon^{ab}f_{ikj}e_{a}{^{k}},\ D=\epsilon^{ab}f_{ikj}\lambda_{a}{^{k}},\ F=\epsilon^{ab}\eta_{ij},\ H=\epsilon^{ab}e_{aj},\ I=\epsilon^{ab}\lambda_{aj},\nonumber
\]
\[
\overline{A}=\left(\partial_{a}\eta_{ij}+f_{ikj}A_{a}{^{k}}\right),\ B=\epsilon^{ab}f_{ijk}f^{k}{_{lm}}e_{a}{^{j}}e_{b}{^{l}}, \ \overline{C}=f_{ikj}e_{a}{^{k}},\ \overline{D}=f_{ikj}\lambda_{a}{^{k}},\ E=\epsilon^{ab}f_{ijk}f^{k}{_{lm}}\lambda_{a}{^{j}}\lambda_{b}{^{m}},\nonumber\\
\]
\[
\overline{F}=\epsilon_{ab}\eta^{ij},\hphantom{11} G=2\theta\mu e_{b}{^{l}}+\lambda_{b}{^{l}}, \hphantom{11} \overline H = \epsilon^{ab}f_{ijk} e_a^je_b^k.\nonumber
\]
The generalized Faddeev-Jackiw bracket $\{,\}_{F-J}$ between two elements of the symplectic variable set (\ref{variables}),  is defined as
\begin{equation}
\{\xi_{i}^{(4)}(x),\xi_{j}^{(4)}(y)\}_{F-J}\equiv\left(f_{ij}^{(4)}\right)^{-1}.
\end{equation}
We thus arrive at the non-vanishing Faddeev-Jackiw brackets for TMG
\begin{eqnarray}
\{A^{i}{_{a}}(x), A^{j}{_{b}}(y)\}_{F-J} &=& \frac{\mu}{2\theta}\eta^{ij}\delta^{2}(x-y),	\\
\{A^{i}{_{a}}(x), \lambda^{j}{_{b}}(y)\}_{F-J} &=& \mu\epsilon_{ab}\eta^{ij}\delta^{2}(x-y), 	\\
\{\lambda^{i}{_{a}}(x), \lambda^{j}{_{b}}(y)\}_{F-J} &=& 2\theta\mu\epsilon_{ab}\eta^{ij}\delta^{2}(x-y),	\\
\{e^{i}{_{a}}(x), \lambda^{j}{_{b}}(y)\}_{F-J} &=& \epsilon_{ab}\eta^{ij}\delta^{2}(x-y).
\end{eqnarray}
 These F-J brackets coincide with the Dirac brackets reported in \cite{blagojevic}. In addition, we can carry out the counting of degrees of freedom as follows. There are $ 18  $ canonical variables $(e^{i}{_{a}}, \lambda^{i}{_{a}}, A^{i}{_{a}})$ and $ 17$ independent constraints $(\Xi_{i}^{(0)},\Theta_{i}^{(0)}, \Sigma_{i}^{(0)}, \Lambda, e^{i}{_{0}}, A^{i}{_{0}}, \varphi)$. Thus, we conclude that 3D TMG has one physical degree of freedom (number of canonical variables $-$ number of independent constraints ), corresponding to the massive graviton, as expected.

\section{ Summary and conclusions}
In this paper,  the dynamical structure of  TMG theory has been studied via the F-J framework. We have obtained  the fundamental gauge structure as well as the physical content of this theory in an  alternative way to that reported in \cite{Mu, blagojevic, Carlip}. It was shown that in the F-J approach is not necessary to classify  the constraints into first- and second-class ones. In this respect,  all the constraints are  treated at the same footing. The correct identification of the constraints  of TMG theory allowed us to show that there is one local physical degree of freedom, and obtain the gauge generators that yield the Poincar\'e symmetries and the diffeomorphisms by mapping the gauge parameters appropriately.   Thereafter, the quantization brackets (F-J brackets) were obtained. Our results coincide with what has been previously obtained via the Dirac approach \cite{blagojevic}. It is worth mentioning that  there is no one-to-one correspondence between the constraints that  we have obtained  via the F-J method and those found via the Dirac formalism  \cite{blagojevic}, though both approaches   yield the same results. Our study suggests that the F-J method turn out to be more economical, unambiguous and straightforward than  Dirac's  one. Finally,  we would like to comment that  according to our results the F-J approach could be useful for studying interesting  features of  models of massive gravity, which include TMG as a particular sector, for instance, topologically massive AdS  gravity. This idea is in progress and will be the subject of forthcoming works \cite{bigravity}.
\newline
\newline
\newline
\noindent \textbf{Acknowledgements}\\[1ex]
This work was supported by CONACyT under Grant No.CB-$2014$-$01/ 240781$. We would like to
thank D. Grumiller for valuable comments and R. Cartas-Fuentevilla for discussion on the subject and reading of the manuscript.
\appendix
\section{Linearized analysis in metric formalism}
In this appendix, using the metric formulation of TMG given by the equation (\ref{second-eq}), we study the linearized theory as a perturbation of the metric about a Minkowski background solution, writing
\begin{equation}
g_{\mu\nu}=\bar{g}_{\mu\nu}+h_{\mu\nu},
\end{equation}
where  $\bar{g}_{\mu\nu}$ is the Minkowski metric and  $h_{\mu\nu}$ is the perturbation. To first-order in this perturbation, the Ricci tensor and the Ricci scalar,  are given by
\begin{eqnarray}
R^{(1)}_{\mu\nu}&=&\frac{1}{2}\left(-\bar{\nabla}^{2}h_{\mu\nu}-\bar{\nabla}_{\mu}\bar{\nabla}_{\nu}h+\bar{\nabla}^{\sigma}\bar{\nabla}_{\nu}h_{\sigma\mu}+\bar{\nabla}^{\sigma}\bar{\nabla}_{\mu}h_{\sigma\nu} \right),\\
R^{(1)}&=&\bar{\nabla}_{\mu}\bar{\nabla}_{\nu}h^{\mu\nu}-\bar{\nabla}^{2}h,
\end{eqnarray}
here $h\equiv\bar{g}^{\mu\nu}h_{\mu\nu}$, and $\bar{\nabla}$ is the covariant derivative constructed with the background metric. Using these expressions one can build the first-order correction of the Einstein, and Cotton tensors as
\begin{eqnarray}
G^{(1)}_{\mu\nu}&=&R^{(1)}_{\mu\nu}-\frac{1}{2}\bar{g}_{\mu\nu}R^{(1)},\\
C^{(1)}_{\mu\nu}&=&\epsilon_{\mu}{^{\alpha\beta}}\bar{\nabla}_{\alpha}\left(R^{(1)}_{\beta\nu}-\frac{1}{4}\bar{g}_{\beta\nu}R^{(1)}\right).\label{cotton_lin}
\end{eqnarray}
On the other hand, the linearized Bianchi identity becomes
\begin{equation}
C^{(1)}_{\mu\nu}-C^{(1)}_{\nu\mu}=0.
\end{equation}
The last term in the right hand side of (\ref{cotton_lin}) is totally antisymmetric on $\mu$ and $\nu$, and therefore merely subtracts the antisymmetric piece from the first term in the right hand side of (\ref{cotton_lin}). We alternatively have
\begin{equation}
C^{(1)}_{\mu\nu}=\frac{1}{2}\left(\epsilon_{\mu}{^{\alpha\beta}}\bar{\nabla}_{\alpha}R^{(1)}_{\beta\nu}+\epsilon_{\nu}{^{\alpha\beta}}\bar{\nabla}_{\alpha}R^{(1)}_{\beta\mu}\right).
\end{equation}
Note also that it is not hard to verify that
\begin{equation}
\bar{\nabla}^{\mu}C^{(1)}_{\mu\nu}=0,\hphantom{111}\mathrm{and}\hphantom{111}  C^{(1)}{^{\mu}_{\mu}}=0.
\end{equation}
Then, the first-order correction of Eq. (\ref{second-eq}), is given by
\begin{equation}
{G}^{(1)}_{\mu\nu}+\frac{1}{\mu}{C}^{(1)}_{\mu\nu}=0.\label{second-lin}
\end{equation}
Furthermore from the trace of this equation one finds that: ${R}^{(1)}=0$, independent of $\mu$. Substituting this back, we therefore find that the Eq. (\ref{second-lin}) can be written as
\begin{equation}
{G}^{(1)}_{\mu\nu}+\frac{1}{\mu}\epsilon_{\mu}{^{\alpha\beta}}\bar{\nabla}_{\alpha}G^{(1)}_{\beta\nu}=0.\label{secon-lin-trace}
\end{equation}
 Now we consider the transverse (divergenceless) and traceless conditions on the Minkowski background as
\begin{equation}
\bar{\nabla}^{\mu}h_{\mu\nu}=0,\hphantom{111}\mathrm{and}\hphantom{111}h^{\mu}{_{\mu}}=0.\label{fix_gauge}
\end{equation}
By making use of these conditions (\ref{fix_gauge}), the equation (\ref{secon-lin-trace}) may be recast into the following form
\begin{equation}
\bar{\nabla}{^{2}}\left(\delta^{\beta}_{\mu}+\frac{1}{\mu}\epsilon_{\mu}{^{\alpha\beta}}\bar{\nabla}_{\alpha}\right)h_{\beta\nu}.
\end{equation}
Furthermore, this equation can be expressed compactly as
\begin{equation}
\left[\mathcal{O}(0)^{2}\mathcal{O}(\mu)h\right]_{\mu\nu}=0,\label{eq_op}
\end{equation}
by introducing two mutually commuting operators as
\begin{equation}
{\mathcal{O}}(0)^{\beta}_{\mu}\equiv\epsilon_{\mu}{^{\alpha\beta}}\bar{\nabla}_{\alpha},\hphantom{111}\mathrm{and}\hphantom{111}\mathcal{O}(\mu)_{\mu}^{\beta}\equiv\delta^{\beta}_{\mu}+\frac{1}{\mu}\epsilon_{\mu}{^{\alpha\beta}}\bar{\nabla}_{\alpha }.
\end{equation}
Since the two operators conmute, the equation (\ref{eq_op}) has two branches of solutions. First, the massive graviton $h^{M}_{\mu\nu}$, given by
\begin{equation}
\left[\mathcal{O}(\mu)h^{M}\right]_{\mu\nu}=h^{M}_{\mu\nu}+\frac{1}{\mu}\epsilon_{\mu}{^{\alpha\beta}}\bar{\nabla}_{\alpha}h^{M}_{\beta\nu}=0.\label{c}
\end{equation}
 The other branch is massless graviton  $\breve{h}_{\mu\nu}$, given by
\begin{equation}
\left[{\mathcal{O}}(0)\breve{h}\right]_{\mu\nu}=\epsilon_{\mu}{^{\alpha\beta}}\bar{\nabla}_{\alpha}\breve{h}_{\beta\nu}=0,
\end{equation}
 which is also solution of Einstein gravity: $G_{\mu\nu}=0$.  Now, let us  define the linear operator $\mathcal{O}(-\mu)_{\mu}^{\beta}\equiv\delta_{\mu}^{\beta}-\frac{1}{\mu}\epsilon_{\mu}{^{\alpha\beta}}\bar{\nabla}_{\alpha}$, which conmute with $\mathcal{O}(\mu)$.
By acting on (\ref{c}) with $\mathcal{O}(-\mu)$, we get the second-order equation for massive graviton
\begin{equation}
\left[\bar{\nabla}^{2}-\mu^{2}\right]h^{M}_{\mu\nu}=0,\label{mass-graviton}
\end{equation}
Similarly, in the massless case, the second-order equation is given by
\begin{equation}
\bar{\nabla}^{2}\breve{h}_{\mu\nu}=0.
\end{equation}
Then, the mass of the massive graviton can be identified by comparing the second-order
equation of motion of massive graviton with that of massless graviton, therefore, the mass of massive graviton is $m=\sqrt{\mu^{2}}$. In addition, the equation  (\ref{c}) propagates a single mode, which has spin-2, because  $h$ is a symmetric traceless second-order tensor, therefore, the equation  (\ref{mass-graviton}) is exactly the Fierz-Pauli equation describing a massive Spin-2  field in Minkowski spacetime.

%%%%%%%%%%%%%%%%%%%%%%%%%%%%%%%%%%%%%%
\section{Gauge symmetry}
We will assume that all the FJ constraints have been identified and therefore only the zero-modes associated with gauge symmetries are still present. In this manner, the final symplectic Lagrangian can be written as  
\begin{equation}
L(\xi)=a_{i}(\xi)\dot{\xi}^{i}+ \dot{\gamma}_a \Omega^a-V(\xi)\hphantom{111}(i=1,2,3,...,N), (a, b= 1,2,...,M), \label{lag}
\end{equation}
here $\Omega^a$'s are the complete set of FJ constraints and  either $\xi$'s or $\gamma$'s form a set of gauge fields.  Now, the symplectic matrix, namely $\bar{f}_{kl}$,   constructed out with the $\xi^i$ variables is not singular, hence $det \bar{f}_{kl}\neq0$. Now let us call to  $f$  the symplectic matrix constructed out by using  the $\xi$'s and $\gamma$'s, that symplectic matrix is singular and will be given by 
\begin{eqnarray}
\label{eq}
&&
f= \left(
 \begin{array}{cc}
 \bar{f}   &  \frac{\partial \Omega}{\partial \xi}   	 	 \\
 -  ( \frac{\partial \Omega}{\partial \xi} )^T&  0    \\
\end{array} 		
\right), 
\label{6x}
\end{eqnarray}
hence  (\ref{6x}) may have M zero-modes of the form 
\begin{eqnarray}
\label{eq}
&&
v^a_k= \left(
 \begin{array}{c}
- (\bar{f})^{-1}  \frac{\partial \Omega}{\partial \xi}   	 	 \\
 1^a  \\
\end{array} 		
\right), 
\label{6c}
\end{eqnarray}
where $(1^a)$ is a   $(M \times1)$ column of zeros except its a-th entry \cite{Mon-Woc}.  
Now, let us assume that the gradient of the potential is orthogonal to all zero-modes, hence, they   must be the generators of the symmetry transformation that leave the action invariant. In this manner, the symmetry of the action over the constraint surface is given by
\begin{eqnarray}
\delta\xi_{i}&=& - (\bar{f})^{-1}  \frac{\partial \Omega} {\partial \xi }\epsilon_{l}, \nonumber \\
\delta{\gamma}_a&=& -\epsilon_I. 
\label{B4}
\end{eqnarray}
here $\epsilon_{l}$ form a set of  infinitesimal parameters that characterize the transformations. It is important to comment that these transformations may reflect either the gauge or reparametrization properties of an invariant theory \cite{Mon-Woc}. In fact, the first relation of the transformations (\ref{B4}) are equivalent one  to  the Dirac gauge transformations obtained from the first class constraints. The latter has not an easy description in the canonical formalism, this can be seen in models such as Floreanini and Jackiw chiral boson and 2D Maxwell fields \cite{F-R}. \\
 Furthermore,  let us finish the appendix   showing  the invariance of the action. In fact, it is well-known, the symmetries are defined by those variation $\delta\xi$ such that the functional variation of the action vanish, this is
\begin{equation}
\delta S=\int dt\left(\frac{\partial L}{\partial \xi_{k}}-\partial_{t}\frac{\partial L}{\partial\dot{\xi}_{k}}\right)\delta\xi_{k}\equiv\int dt \left(f_{km}\dot{\xi}^{m}-\frac{\partial V}{\partial\xi}_{k}\right)\delta\xi_{k}=0.\label{vari}
\end{equation}
Therefore, this expression defines the gauge symmetry; if there exists some variation $\delta\xi_{k}$ satisfying the Eq. (\ref{vari}), then the transformation
\begin{equation}
\xi_{k}\longrightarrow\xi_{k}+\delta\xi_{k},
\end{equation}
is a symmetry of the action S. Hence, we can construct a variation $\delta\xi_{k}$ satisfying (\ref{vari}) on the constraints surface, given as
\begin{equation}
\delta\xi_{k}=\left(v_{l}\right)_{k}\epsilon^{l},
\end{equation}
Therefore, since $(v^{l})_{i}$ are the zero-modes on the constraints surface, they must satisfy the equation of motion, i.e.
\begin{equation}
\int dt\left(f_{km}\dot{\xi}_{m}-\frac{\partial V}{\partial\xi_{k}}\right)(v^{l})_{k}\epsilon_{l}=\int dt \epsilon_{l}(v^{l})_{k}^{T}\left(f_{km}\dot{\xi}_{m}-\frac{\partial V}{\partial\xi_{k}}\right)=\epsilon_{l}\Omega^{(l)}.
\end{equation}
And this shows that the action is invariant under displacements in directions orthogonal to the gradient of the potential.

\end{document}